# Computational Aerothermal Framework and Analysis of Stetson's Mach 6 Blunt Cone

**Arturo Rodriguez** [1], **Piyush Kumar** [1], **Cesar Diaz-Caraveo** [1], **Richard O. Adansi** [1], **Luis F. Rodriguez**[2], **Vineeth V. K**[3] **and Vinod Kumar** [3, *]

1    Aerospace and Mechanical Engineering Department, University of Texas at El Paso, El Paso 79968, Texas

2    Mechanical and Aerospace Engineering, Clarkson University, Potsdam 13699, New York

3    Department of Mechanical and Industrial Engineering, Texas A&M University at Kingsville, Kingsville 78363, Texas

*    Correspondence: vinod.kumar@tamuk.edu

**Abstract:** Accurately predicting aerothermal behavior is paramount for the effective design of hypersonic vehicles, as aerodynamic heating plays a pivotal role in influencing performance metrics and structural integrity. This study introduces a computational aerothermal framework and analyzes a blunt cone subjected to Mach 6 conditions, drawing inspiration from Stetson's foundational experimental work published in 1983. While the findings offer significant insights into the phenomena at play, the study highlights an urgent necessity for integrating chemical kinetics to comprehensively capture non-equilibrium effects, thereby enhancing the predictive accuracy of computational fluid dynamics (CFD) simulations. This research implements a one-way coupling method between CFD simulations and heat conduction analysis, facilitating a thorough investigation of surface heat transfer characteristics. The numerical results elucidate discrete roughness elements' impact on surface heating and fluid dynamics within high-speed airflow. Furthermore, the investigation underscores the critical importance of accounting for non-equilibrium thermochemical effects in aerothermal modeling to bolster the accuracy of high-enthalpy flow simulations. By refining predictive computational tools and deepening understanding of hypersonic aerothermal mechanisms, this research lays a robust groundwork for future experimental and computational endeavors, significantly contributing to advancing high-speed flight applications.

## 1. Introduction

Accurately predicting aerothermal behavior in hypersonic flows represents a mission-critical endeavor in designing and operating high-speed aerospace vehicles. Aerodynamic heating significantly influences surface temperatures, structural integrity, and overall vehicle performance, rendering precise modeling essential for defense and space applications. Traditional empirical correlations, particularly those predicated on the Reynolds number, frequently fail to capture the complexities inherent in hypersonic aerothermal phenomena [1,2]. Computational Fluid Dynamics (CFD) has become a prevalent tool for investigating hypersonic flow behavior; however, several limitations often constrain its application in aerothermal modeling. Current computational methodologies struggle to incorporate complex flow interactions, including surface roughness, shock interactions, and high-temperature gas effects [3]. The critical nature of this challenge emphasizes the pressing necessity for advancing numerical techniques tailored for hypersonic simulations, thereby underscoring the importance of integrating higher-fidelity physical models to enhance predictive accuracy. Despite the significant computational demands associated with these advanced techniques, integrating robust turbulence modeling approaches, such as Large-Eddy Simulations (LES) and Direct Numerical Simulations (DNS), proves indispensable. Furthermore, implementing finite-rate chemistry models is crucial for addressing thermochemical non-equilibrium effects, which play a pivotal role in high-enthalpy hypersonic flows [4].

Understanding the behavior of hypersonic vehicles necessitates a comprehensive investigation of aerothermal and structural effects [5,6]. A detailed analysis of aerodynamic heating and its ramifications on structural deformation can be achieved by integrating computational fluid dynamics (CFD) with computational thermal and structural dynamics (CTSD). This coupled approach accentuates the criticality of incorporating multi-physics interactions in simulations pertinent to hypersonic flight. Despite substantial advancements in these methodologies, considerable challenges remain in accurately modeling the effects of surface roughness on thermal loads and material responses [7,8]. To address these challenges, the present study aims to formulate a novel computational aerothermal framework that scrutinizes how surface roughness influences heating and material response for a blunt cone traveling at Mach 6. By implementing a one-way coupling between CFD-generated surface temperatures and solid-state heat transfer simulations, this methodology facilitates a more nuanced evaluation of aerothermal loads attributable to surface roughness [9]. This research builds upon foundational experimental studies and seeks to enhance predictive capabilities concerning high-speed heating effects by integrating advanced thermochemical models. Such an approach is especially pertinent in light of the burgeoning interest in hypersonic technologies. The accurate thermal modeling afforded by this investigation has the potential to inform material selection and vehicle design, thereby optimizing performance and thermal protection measures, underscoring the practical significance of this study in the context of hypersonic advancements [10–12].

This study aims to bridge the gap between empirical and computational methodologies in hypersonic aerothermal modeling. The advanced simulation framework presented here is designed to significantly enhance predictive capabilities, addressing the limitations of current computational techniques. This innovation is poised to greatly improve the accuracy of aerothermal predictions, thereby advancing the design of hypersonic vehicles. The implications of this research are particularly significant for vehicle design, as the improved predictive accuracy may lead to more efficient and safer hypersonic flight operations. Accurately predicting aerothermal loads is crucial in hypersonic flight, given their substantial influence on vehicle design and material selection. The foundational experiments conducted by Stetson in 1983, which provided critical data on boundary layer behavior over blunt cones at elevated Mach numbers, have been a cornerstone in this field [9]. Recent computational studies have built upon this work, seeking to replicate and extend these findings by applying advanced numerical methods [13].

A notable advancement in hypersonic flow modeling is the introduction of chemical kinetics computational fluid dynamics (CFD) solvers, which effectively account for the non-equilibrium dynamics between translational-rotational and vibrational-electronic energy modes. Casseau et al. have developed 'hy2Foam,' a pivotal open-source chemical kinetics CFD solver integrated within the OpenFOAM framework, specifically engineered to predict hypersonic reacting flows [14]. This solver can discern between translational-rotational and multiple vibrational-electronic temperature states, thereby facilitating the modeling of vibrational-translational and vibrational-vibrational energy exchanges within an eleven-species air mixture [15,16]. Implementing such advanced models is vital for accurately simulating the thermochemical non-equilibrium phenomena characteristic of hypersonic flows, and the 'hy2Foam' solver is a significant step forward in this field.

Integrating computational fluid dynamics (CFD) with heat conduction analysis facilitates a comprehensive investigation of surface heat transfer characteristics. This synergistic approach advances our understanding of discrete roughness elements that influence surface heating and fluid dynamics within high-speed flow regimes. By honing computational predictive tools, this research plays a significant role in the progression of hypersonic aerothermal mechanisms. It establishes a foundational framework for subsequent experimental and computational inquiries, thereby promoting the development of applications related to high-speed flight. The transition from Stetson's seminal experimental observations to contemporary computational analyses underscores the necessity of incorporating sophisticated modeling techniques, such as chemical kinetics models, to effectively capture the intricate non-equilibrium phenomena intrinsic to hypersonic aerothermal research [17–19]. This integration is essential for developing predictive tools to inform next-generation hypersonic vehicles' design and optimization [20].

## 2. Applications

The computational aerothermal framework and analysis, a unique and significant contribution of this research, have several critical applications in the field of hypersonic vehicle design and operation:

1. **Thermal Protection System (TPS) Design**: The computational aerothermal framework is pivotal in enhancing (TPS). It provides precise and reliable predictions of surface heat flux, which is a fundamental factor in designing effective TPS. These systems protect vehicles from the extreme aerodynamic heating encountered during hypersonic flight. The framework's capability to evaluate heat loads is instrumental in selecting appropriate materials and configurations, thereby ensuring hypersonic vehicles' structural integrity and operational safety [20].

2. **Aerodynamic Performance Optimization**: Optimizing aerodynamic performance is critically informed by a comprehensive understanding of aerothermal effects. This understanding facilitates the refinement of vehicle geometries aimed at minimizing aerodynamic drag and effectively managing thermal distribution. Empirical studies have demonstrated that the geometry of leading-edge bluntness can substantially influence heat transfer dynamics and aerodynamic characteristics. For instance, research indicates that blunting the leading edge of a vehicle can reduce heat transfer rates by increasing the bow shock standoff distance, thereby decreasing aerodynamic heating. However, this modification may also lead to a decrease in aerodynamic performance due to increased drag [20,21]. Therefore, strategic design decisions must balance the benefits of reduced thermal loads against potential aerodynamic penalties.

3. **Structural Integrity Assessment**: An Integrated Approach. The Analysis of structural integrity in hypersonic flight is a complex task that necessitates a comprehensive understanding of the interplay between aerothermal loads and structural dynamics. By employing a coupled approach, engineers are empowered to predict areas susceptible to deformation or potential failure. This responsibility ensures the vehicle's structural framework is robust enough to endure the extreme stresses experienced during hypersonic operation. This integrative methodology, driven by the expertise and dedication of engineers, is paramount for sustaining aerospace vehicles' safety and operational functionality under the rigorous conditions associated with hypersonic travel.

4. **Material Selection and Testing**: The framework established for material selection and testing is instrumental in comprehensively evaluating materials subjected to simulated hypersonic conditions. That facilitates a systematic approach to identifying and selecting suitable composites, ceramics, or metals that can withstand extreme temperatures and significant mechanical stresses. Recent scholarly research highlights the necessity of designing materials specifically for hypersonic applications, with a pronounced focus on critical factors such as thermal resistance and structural integrity [22]. Such a targeted approach underscores the importance of advancing material science to meet the rigorous demands of hypersonic flight.

5. **Validation of Computational Models**: The insights derived from this framework have significant implications for validating and enhancing computational fluid dynamics (CFD) models, thereby improving their predictive accuracy concerning hypersonic flows. Such validation is imperative for developing reliable simulations that guide both design and operational decision-making processes [23].

By exploring these critical areas, the present research makes a unique and significant contribution to the field of hypersonic technology, thereby facilitating the development of vehicles designed for sustained high-speed flight, emphasizing optimized performance and enhanced safety measures.

The computational aerothermal framework and analytical methodology established in this study, primarily focused on Stetson's Mach 6 blunt cone, possess several specific applications:

1. **Validation of Computational Methods**: The present study establishes a benchmark for validating computational fluid dynamics (CFD) and heat conduction simulations under hypersonic flow conditions. By conducting a comparative analysis between numerical results and the experimental data obtained from Stetson, researchers can rigorously evaluate the accuracy of their models in predicting critical phenomena such as surface heat transfer and boundary-layer behavior. This validation process is essential for advancing the reliability and applicability of computational methodologies in hypersonic aerothermodynamics and has significant implications for the design and performance of hypersonic vehicles.

2. **Design of Hypersonic Test Articles**: The design of hypersonic test articles is critically informed by insights derived from analyses of surface roughness and heat transfer effects on blunt cone geometries. A comprehensive understanding of these factors is essential for engineers to create accurate models that effectively simulate flight conditions. Their role in enhancing the reliability of testing methodologies and the quality of data collected during hypersonic wind tunnel experiments is crucial. Such advancements are vital for developing hypersonic technologies and their applications [20].

3. **The Development of Transition Prediction Tools**: This framework facilitates the creation of predictive tools for boundary-layer transition induced by surface roughness in hypersonic flow regimes. The precise forecasting of transition phenomena is essential for optimizing the design of thermal protection systems and enhancing hypersonic vehicles' overall performance [24,25].

4. **Enhancement of Aerothermal Databases**: This study significantly contributes to existing aerothermal databases by supplying comprehensive computational data regarding heat transfer and boundary-layer characteristics over blunt cones at a Mach number of 6. Such detailed information is essential for validating and refining these empirical correlations, which is crucial as it aids in designing hypersonic vehicles, thereby advancing the field of aerospace engineering [26].

By systematically addressing these focal areas, the research contributes to a more profound understanding of hypersonic aerothermal phenomena. It facilitates the advancement of design and analytical methodologies for high-speed aerospace vehicles.

## 3. Method

This research employs a one-way coupling computational methodology integrating steady-state Navier-Stokes-based Computational Fluid Dynamics (CFD) simulations with a heat conduction solver. This robust framework facilitates a rigorous analysis of the influence of surface roughness on aerothermal characteristics for a blunt cone at Mach 6. The study's findings have practical implications for the design and performance of high-speed vehicles. By enabling an in-depth investigation of surface temperature distribution and its implications for roughness-induced heating, it advances the understanding of heat transfer phenomena in high-speed fluid dynamics. The employed approach is validated against the experimental heat transfer coefficient outlined by Stetson, thereby ensuring its physical accuracy and reliability.

### 3.1 Computational Framework

The numerical approach comprises two distinct stages of investigation:

1. Computational Fluid Dynamics (CFD) Simulation of the External Flow Field

   a. This phase involves applying a steady-state, axisymmetric, compressible, viscous Navier-Stokes solver that models the laminar hypersonic flow across a blunt cone configuration.
   b. The Navier-Stokes solver plays a key role in this phase, as it computes and elucidates the surface temperature and heat flux distributions at the cone's exterior.
   c. A no-slip boundary condition is imposed at the wall interface to ensure the fidelity of the temperature predictions. This condition is crucial as it prevents the fluid from slipping past the wall, which could lead to inaccurate temperature predictions.
   d. Additionally, discrete roughness elements are introduced along the height of the cone's outer wall to account for real-world surface characteristics.

2. Heat Conduction Simulation in the Solid Domain

   a. After the CFD analysis, the surface temperature data obtained from the Navier-Stokes solver is utilized as a boundary condition for the subsequent heat conduction simulation.

b.  A commercial finite element heat conduction solver is employed to thoroughly resolve the heat diffusion equation within the solid domain of the stainless-steel cone, ensuring the validity of the results.

c.  Furthermore, the heat transfer coefficient is computed to facilitate validation of the numerical heat flux predictions against empirical measurements, thereby ensuring the results' robustness.

### 3.2 Analysis Flowchart for One-Way Coupling Method

To elucidate the computational workflow associated with our proposed methodology, we present an analysis flowchart in Figure 1. This flowchart delineates the sequential steps involved in implementing our one-way coupling approach, thereby enhancing the clarity of the execution process.

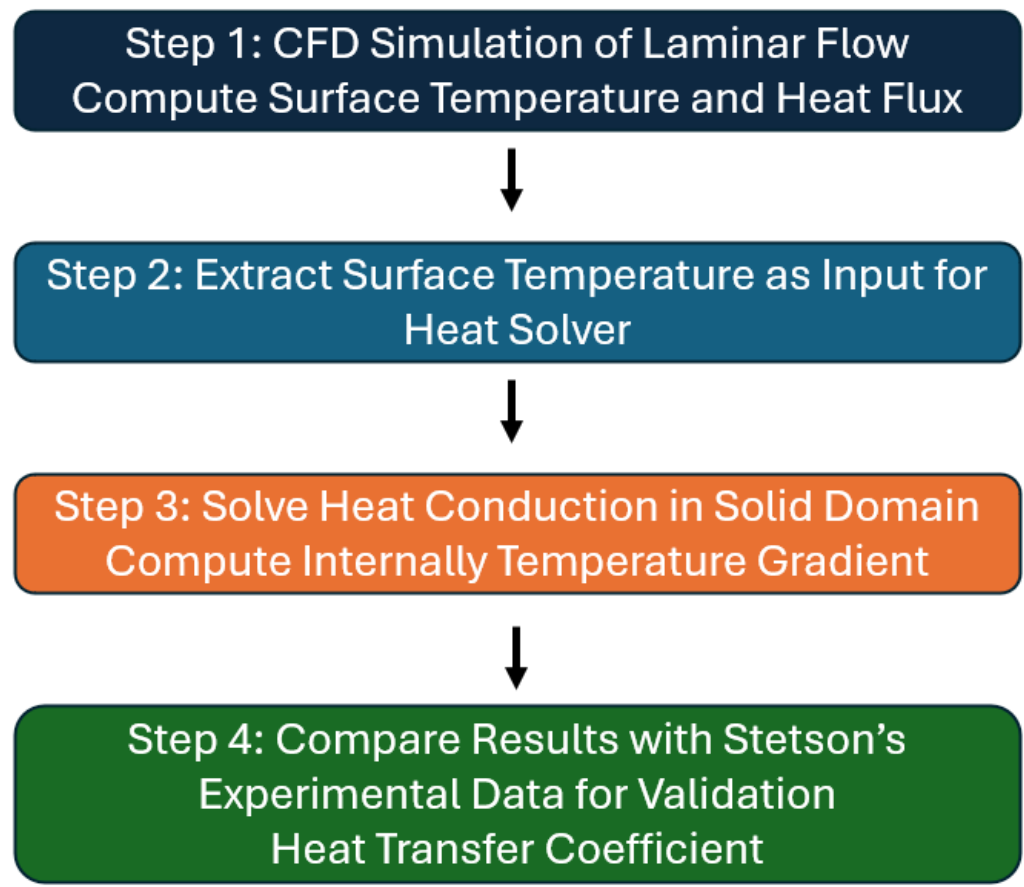

**Figure 1.** Aerothermal Analysis Schematic

### 3.3 Physical Validity of the Proposed Computational Technique

Well-Established Navier-Stokes and Heat Conduction Solvers:

**ANSYS Fluent Validation:** ANSYS Fluent has undergone extensive validation to simulate hypersonic flows. A notable study at Washington University in St. Louis successfully modeled hypersonic flow over a yawed cone using ANSYS Fluent, demonstrating its efficacy in accurately capturing intricate flow phenomena at elevated Mach numbers [27]. This validation not only underscores the solver's robustness in addressing complex aerodynamic challenges within hypersonic regimes but also directly supports the physical validity of the proposed computational techniques.

**ANSYS Transient Thermal Analysis Solver:** The Transient Thermal Analysis solver integrated within ANSYS Mechanical is founded upon rigorous thermal modeling principles, which are of utmost importance in ensuring precise simulation of heat transfer processes within solid domains. The solver's reliance on these established thermal dynamics principles contributes to its credibility, as corroborated by comprehensive documentation from ANSYS regarding its capabilities in performing time-dependent thermal analyses with high accuracy and reliability.

Comparison with Experimental Data (Stetson's Study): Our work is part of a larger research community, and we contribute to it by comparing our results with those of Stetson's Study. This comparison helps us validate our computational methodology and ensures that we are on the same page with the broader research community.

- **Heat Transfer Validation:** The validation of our computational methodology is substantiated through a systematic comparison with experimental heat flux data, thereby reinforcing the credibility of our heat transfer results.

Justification for One-Way Coupling Approach:

- **Computational Efficiency:** Our one-way coupling technique, in contrast to fully coupled aerothermal models, is highly efficient in capturing the principal transition mechanisms in the laminar regime. This practical and cost-effective approach allows us to achieve our research objectives with significantly fewer computational resources.
- **Established Reliability:** The one-way coupling approach, as demonstrated in several hypersonic boundary-layer transition studies, has proven its reliability. For instance, a notable investigation into hypersonic flow over a yawed cone employed a comparable methodology, highlighting the efficacy of one-way coupling in accurately representing essential flow physics. This established reliability should instill confidence in its use for our heat transfer analysis.

### 3.4 Equations Solved and Set-Up

In this study, we chose to employ a compressible steady-state viscous Navier-Stokes solver due to its proven effectiveness in addressing similar problems in the past. This solver, which has been widely used in the field of fluid dynamics, was selected for its ability to accurately model the behavior of compressible fluids under steady-state conditions [28,29].

The governing equations we solve include the conservation of mass, momentum, and energy.

$$\frac{\partial}{\partial x_j}(\rho u_j) = 0 \tag{1}$$

$$\frac{\partial}{\partial x_j}(\rho u_i u_j) = -\frac{\partial p}{\partial x_i} + \frac{\partial}{\partial x_j}(\tau_{ij}), i = 1,2 \tag{2}$$

$$\frac{\partial}{\partial t}\left(\rho e + \frac{1}{2}\rho u^2\right) + \frac{\partial}{\partial x_j}\left[\left(\rho e + \frac{1}{2}\rho u^2\right)v_j\right] = -\frac{\partial}{\partial x_j}(pu_j) + \frac{\partial}{\partial x_j}(\tau_{ij}u_i) + \frac{\partial}{\partial x_j}(\dot{q}_j) \tag{3}$$

In our research, we have employed a particular equation formulation, informed by physical experiments, to approximate the desired solution. To accurately depict hypersonic physics, we must adjust the energy equation to encompass classical and quantum effects within the chemical non-equilibrium, such as chemical kinetics models. Employing the finite volume spectral method, we have utilized this formulation with an implicit steady-state solver and the AUSM flux type to solve the problem. It should be noted that this approach only encompasses the physics of the fluid domain [15].

Our approach is grounded in a specific formulation derived from empirical physical experiments, allowing us to approximate the desired solution comprehensively. This rigorous methodology ensures the robustness and validity of our findings. To accurately characterize hypersonic phenomena, it is essential to modify the energy equation to account for classical and quantum effects within a chemically non-equilibrium framework, utilizing chemical kinetic models. We have implemented this formulation through the finite volume spectral method, employing an implicit steady-state solver. The adaptability of the AUSM (Advection Upstream Splitting Method) flux type is instrumental in facilitating the resolution of the problem. It is important to emphasize that this methodology encompasses only the physics of the fluid domain [15].

In the context of thermal analysis, the heat equation within the domain of heat conduction is pivotal.

$$-\nabla \cdot k \nabla T = 0$$

(4)

However, the surface boundary condition truly adds depth and contours to our understanding. It plays a critical role in addressing the solver's transient dynamics, emphasizing spatial factors exclusively. We employ the finite element method to tackle the associated Partial Differential Equation (PDE), a testament to the intricate nature of our research.

This study integrates a Navier-Stokes solver with a heat conduction solver, facilitating a direct comparison with empirical data. The heat transfer coefficient emerges as the central parameter for this comparison and serves as the primary variable of interest. This methodology, with its strong focus on validation against empirical data, ensures the reliability and applicability of our findings. We utilize the surface temperature derived from the Navier-Stokes solver to establish a boundary condition for the heat conduction solver within the solid domain. Subsequently, the heat transfer coefficient is computed based on these findings and benchmarked against the experimental observations documented in Stetson's research. This coupling of solvers creates a coherent framework for validating numerical simulations about actual physical experimental data.

### 3.5 Roughness Model | Modified Law-of-the-Wall

The Modified Law-of-the-Wall framework describes how surface roughness affects the velocity profile in turbulent boundary layers [30,31]. The roughness function $\Delta U^+$ quantifies the velocity shift due to roughness and is typically expressed as a function of the roughness Reynolds number ($Re_k$), which depends on the friction velocity ($u_*$), the roughness height ($k$), and the kinematic viscosity ($\nu$). The friction velocity ($u_*$) is determined using the skin friction coefficient ($C_f$) and the free-stream velocity ($U$), requiring adjustments based on roughness correlations. The velocity profile is then modified as $U^+ = \frac{1}{\kappa} ln(y^+) + B - \Delta U^+$, where $U^+$ is the non-dimensional velocity, $y^+$ is the wall-normal coordinate, $\kappa$ is the von Kármán constant, and $B$ is the intercept for smooth walls. This formulation accounts for adverse and favorable pressure gradients and compressibility effects, which influence turbulence development and wall shear stress. This framework enhances hypersonic flow predictions by integrating roughness parameters into boundary-layer modeling, enabling more accurate aerothermal assessments for high-speed vehicle design.

## 4. Computational Set-Up and Results

In our study, we meticulously constructed a computational model by employing a Navier-Stokes solver to simulate the fluid flow around the blunt cone. We utilized a heat equation commercial solver to model heat conduction. The blue section represents the fluid flow domain, while the dark gray depicts the heat conduction domain in Figure 5. The dimensions of the domains are as follows: The base radius is 2.0 inches, and the nose tip base has a measurement of 0.6 inches with a half angle of 8 degrees. These specific dimensions pertain to the blunt cone utilized by Stetson in his physical experiments at the Wright-Patterson Air Force Base [1].

The purpose of using this cone was to observe how roughness elements can induce transition due to discontinuities on the surface, thus causing disturbances in fluid flow. The varying heights of the roughness elements lead to diverse fluid flow development and the formation of distinct environments, which have practical implications for understanding and predicting fluid flow in real-world scenarios [32–35]. As a consequence of these discontinuities, separation occurs, triggering the transition of the boundary layer. This process results in the formation of horseshoe vortices that are pushed upstream as the fluid flow stagnates, ultimately creating signatures of small shock waves during the development of the fluid flow. Thus, it represents an unsteady time of the simulation, a significant finding with practical implications in fluid dynamics and aerodynamics.

In this study, we conducted a comprehensive axisymmetric simulation that emulates the physical testing conditions of the Stetson Mach 6. The fluid flow was initialized at a Mach number of 6 to achieve a steady state. The simulation parameters were set to reflect the conditions prevalent at an altitude of 4000 feet, with static pressure maintained at 1827.71 lbs/ft² and static temperature at 504.1 R. The boundary condition applied at the vehicle's surface was a no-slip condition consistent with standard practices in fluid dynamics. The vehicle's wall is modeled as composed of stainless steel 17-4 PH, integrating variable conditions within the heat conduction domain. It is important to note that the thermal conductivity and specific heat capacities of the material are considered to be functions of temperature, thereby allowing for a more accurate representation of the thermal response under the given flow conditions.

$$k = 2.08e - 4 + 1.13e - 7 \cdot T \tag{5}$$
$$Cp = 0.104 + 3.38e - 5 \cdot T + 4.45e - 8 \cdot T^2 \tag{6}$$

Equations (5) and (6) represent mathematical models derived from empirical data. They encapsulate the temperature-dependent characteristics of air's thermal conductivity and specific heat capacity. These polynomial formulations are based on experimental measurements to model heat transfer phenomena in high-temperature environments accurately.

The density ($\rho$) of the material under consideration is measured at 0.282 Lbm/in³. An essential and significant system feature is an inner adiabatic wall, which establishes a zero-flux Neumann boundary condition, a key element in our analysis. The wall thickness along the z-axis is specified to be 1 inch. These parameters are pertinent given that the simulation operates under laminar flow conditions, wherein the three components are exclusively characterized by laminar rather than turbulent fluid dynamics. The justification for the amalgamation of two-dimensional modeling with laminar flow assumptions lies in the specific conditions of this analysis.

The present investigation employs boundary conditions within a computational fluid dynamics (CFD) simulation of a compressible viscous flow domain characterized by an axisymmetric configuration. Specific far-field pressure conditions are implemented at the leading boundaries to represent compressible fluid behavior accurately. Furthermore, the simulation adheres to the ideal gas law, treating the fluid as perfect, with an atmospheric pressure outlet established at 4000 feet. Notably, the conical front surface incorporates essential no-slip stationary wall conditions, which are not just pivotal, but absolutely crucial for ensuring the fidelity of the flow representation in this study.

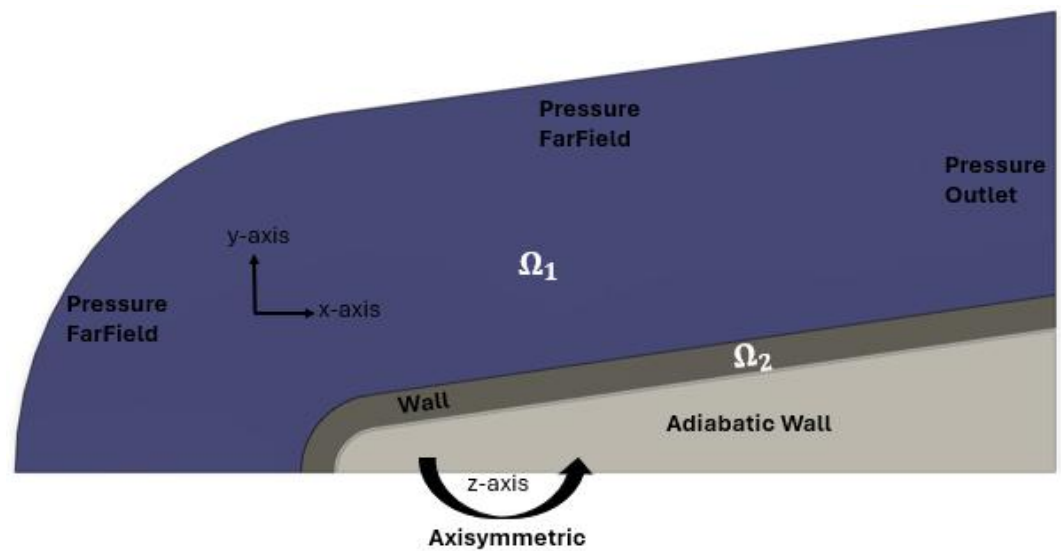

**Figure 2.** Computational Set-Up

The mesh demonstrates low skewness, high orthogonality, and superior element quality, particularly for non-deformed orthogonal structured elements utilized in our simulation. The aspect ratio, a critical factor, must be maintained at four or less to ensure accurate numerical convergence and stability. Our analysis primarily concentrates on the vehicle noses, as this region significantly influences fluid flow dynamics.

This study emphasizes the careful consideration of mesh refinement in computational studies, a crucial aspect of achieving reliable and consistent results. Our analysis examined three distinct mesh configurations: mesh 1 (coarse), mesh 2 (base), and mesh 3 (fine), as outlined in the accompanying table. By stabilizing the drag force, we observed an apparent convergence associated with the variation in element size pertinent to the case under investigation. This finding underscores the importance of drag force as a pivotal parameter for attaining mesh independence.

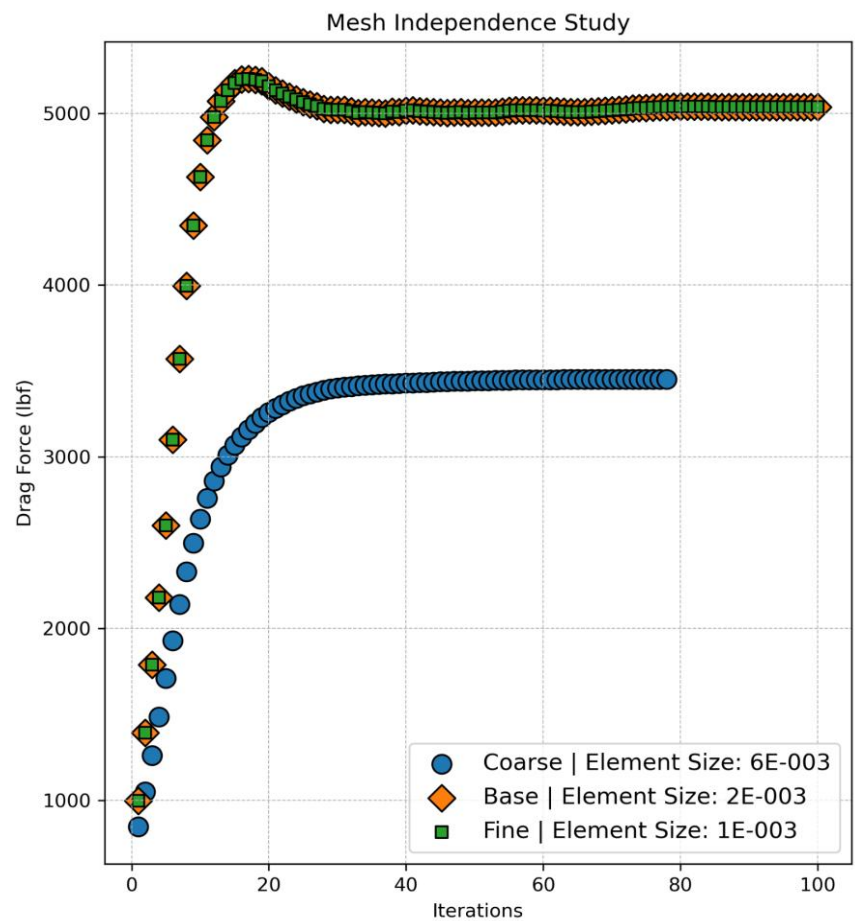

**Figure 3.** Mesh Independence Study

The analysis presented herein pertains to the thermodynamic behavior of air modeled as an ideal gas, conducted under standard atmospheric conditions at 4000 feet. The methodology is anchored in an implicit formulation that employs the Advection Upstream Splitting Method (AUSM) flux type—an essential technical consideration. Spatial discretization is achieved through a least squares cell-based gradient approach, coupled with a second-order upwind scheme for the flow representation, thereby ensuring high accuracy in the computations. With its novel approach, this research promises to unveil new insights into the thermodynamic behavior of air.

A Courant number of 5 was employed throughout the simulation, and the resultant output corresponds to a heat transfer boundary condition tailored explicitly for application to a 17-4 PH stainless steel material. This study's findings, including detailed contour results that elucidate the flow characteristics surrounding a blunt cone geometry, are not just theoretical. They have practical implications, contributing to the broader understanding of aerodynamic behavior in similar contexts and potentially informing real-world applications.

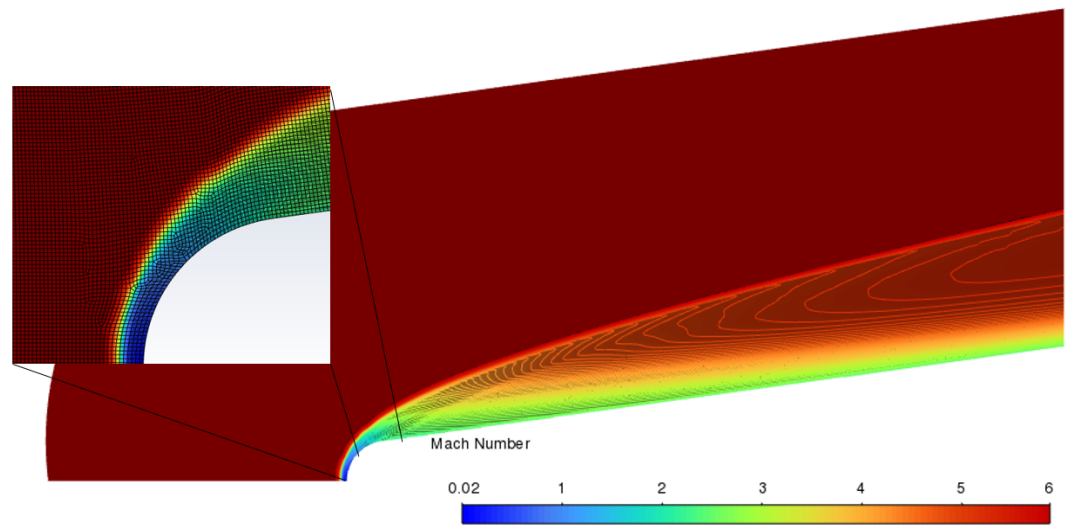

**Figure 4.** Blunt Cone | Laminar Solution | Mach Number Contour

Figure 4 illustrates the Mach number contour corresponding to the laminar flow solution of a blunt cone subjected to hypersonic conditions at Mach 6. This visual representation emphasizes the structured grid employed to effectively resolve the aerothermal characteristics of the flow regime surrounding the cone. The mesh density is notably enhanced in proximity to the vehicle's surface, particularly within the boundary-layer region, to represent shock wave formation and viscous phenomena accurately. The numerical methodology utilized in this analysis is recognized for its robustness, characterized by low skewness, high orthogonality, and superior element quality, all essential for achieving reliable convergence in computational simulations. The development of the shock wave is prominently observed at the leading edge of the cone, subsequently propagating downstream and significantly influencing the pressure and temperature distributions along the vehicle's surface, thereby demonstrating the profound impact of the hypersonic flow dynamics on the vehicle's performance.

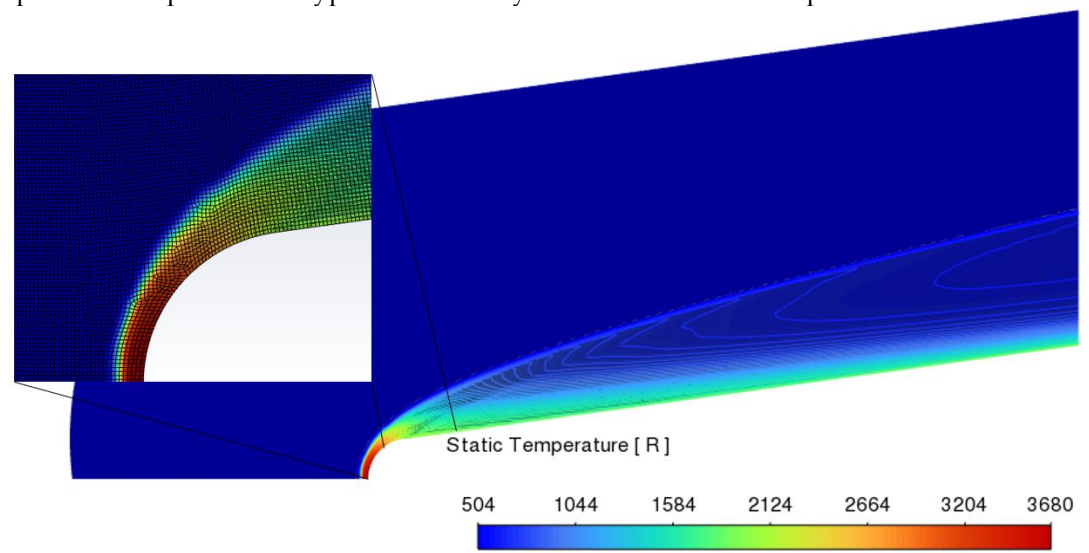

**Figure 5.** Blunt Cone | Laminar Solution | Temperature Contour

Figure 5 depicts the distribution of wall static temperature over the blunt cone under hypersonic flow conditions. The color gradient illustrates the temperature variations, with the highest temperatures concentrated at the stagnation point due to substantial aerodynamic heating. As the flow progresses downstream, the temperature gradually decreases along the cone's frustum. The simulation incorporates the effects of heat conduction within the solid structure, revealing a thermal gradient that extends into the cone's interior. This careful consideration of heat conduction effects further corroborates the expected heat transfer trends in hypersonic aerothermal environments. It enhances the analysis of boundary-layer transitions influenced by roughness-induced perturbations.

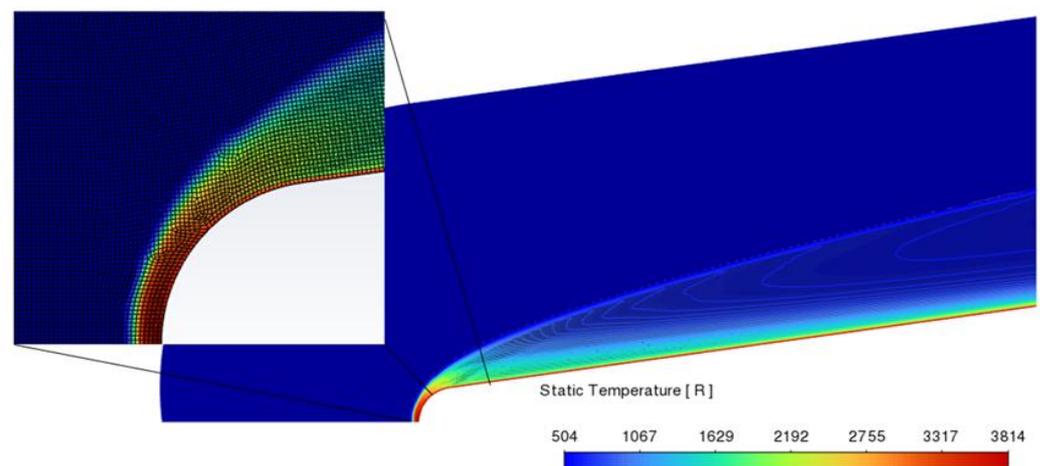

**Figure 6.** Blunt Cone | Turbulent with Roughness Solution | Temperature Contour

Figure 6 presents a detailed illustration of the temperature distribution across the surface of a blunt cone subjected to hypersonic turbulent flow conditions at a Mach number of 6. Where we use the k-omega RANS model to handle the turbulence. The computational analysis, a testament to the complexity of our research, elucidates the effects of surface roughness on thermal loading, revealing significant localized heating variations attributed to disturbances in the boundary layer induced by surface irregularities. The regions of maximum temperature are predominantly located at the stagnation point, where the incident flow experiences substantial compression and subsequent aerodynamic heating. Progressing downstream along the cone's surface, a reduction in the temperature gradient is observed, which is influenced by the development of the boundary layer and the associated transition phenomena.

Introducing roughness elements within fluid dynamics frameworks results in localized temperature spikes, which indicate the early onset of transition and heightened heat transfer rates attributed to augmented turbulence. The interaction between the shock-wave structure and the boundary layer significantly influences the thermal boundary layer thickness and heat flux distribution. This simulation underscores the critical importance of precise roughness modeling in hypersonic aerothermal analyses, as it is essential for accurately predicting boundary-layer transition phenomena and associated heating effects. The urgency of this need is further highlighted by the call for subsequent research endeavors to incorporate high-fidelity turbulence models alongside experimental validation, enhancing the predictive accuracy of computational aerothermal frameworks.

In the context of heat conduction analysis, the formulation of precise boundary conditions is paramount when employing the finite element method (FEM). This methodology leverages basis functions defined between nodes to approximate the solution to the governing equations. Our approach to the problem was grounded in the weak formulation of the partial differential equation, which encapsulates the variational integral form of the equation. The initial conditions were informed by constant temperature boundaries derived from computational fluid dynamics (CFD) simulations. Additionally, we imposed axisymmetric conditions complemented by zero Neumann boundary constraints, ensuring a comprehensive framework for the analysis.

In the context of our collaborative study, we implemented an integrated approach that combines fluid dynamics and the Navier-Stokes equations with principles of heat conduction, explicitly utilizing the Heat Equation solver. This methodology, developed through our collective expertise, enabled us to derive comprehensive aerothermal results, which we quantified through the heat transfer coefficient. Our computational analysis builds upon and extends the foundational work presented by Stetson in his seminal 1983 paper at the Fluid and Plasma AIAA conference. The findings from our investigation directly apply to the physical experiments delineated in Stetson's original research, thereby contributing to a deeper understanding of the subject matter.

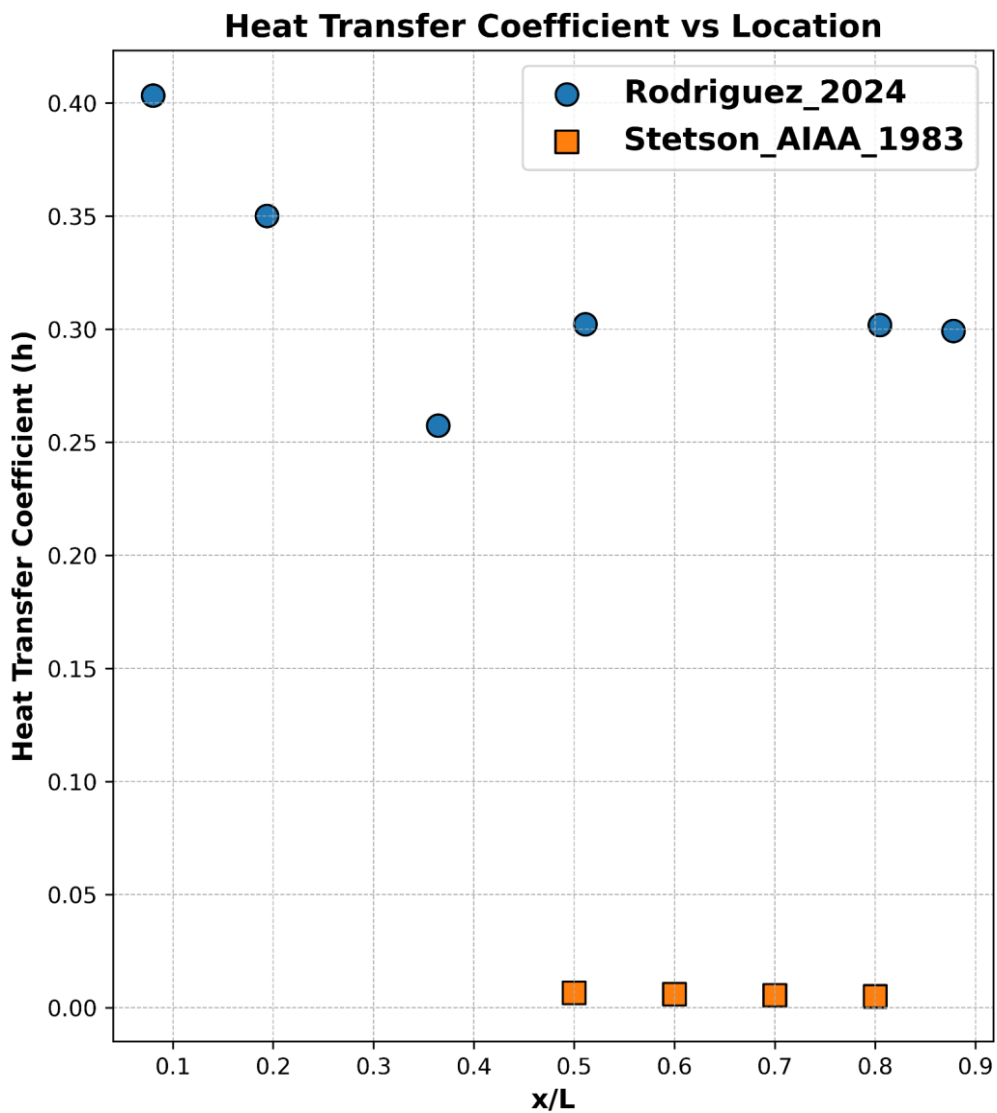

**Figure 7.** Blunt Cone at Mach 6 | Laminar Fluid Flow

The observed discrepancies in the heat transfer coefficient results between Rodriguez (2024) and Stetson (AIAA, 1983) can be attributed to the absence of a more sophisticated energy equation model and chemical kinetics. This presents an exciting opportunity for advancement in our field. Prior literature, including Schneider's comprehensive review, emphasizes that accurately modeling high-enthalpy hypersonic flows necessitates energy equations that effectively account for non-equilibrium effects. These effects, encompassing vibrational relaxation and dissociation processes, are critical in elucidating such flow regimes' intricate energy exchange mechanisms. The omission of these phenomena in the simulations compromises their fidelity, thereby hindering a complete resolution of the aerothermal behavior under extreme flow conditions.

Stetson's investigations elucidate the pivotal function of shock-induced vibrational excitation in the modulation of boundary-layer development and its subsequent influence on heat transfer characteristics. In parallel, Schneider's comprehensive review delineates that the discrepancies between theoretical predictions and experimental findings often arise from insufficient modeling of non-equilibrium processes. Notably, the omission of a chemical kinetics framework in contemporary computational approaches is a significant limitation, as it constrains their capacity to encapsulate critical mechanisms, particularly the interactions between vibrational and electronic energy modes and the associated dissociation processes. These factors significantly affect the boundary layer's evolution and instabilities' growth.

Future research will mitigate the existing limitations by integrating a chemical kinetics models to elucidate the exchanges between vibrational and electronic energy and the dissociation coupling terms. These enhancements are anticipated to foster improved concordance between computational outcomes and experimental data by effectively capturing the complex interactions between relaxation processes and boundary-layer instabilities. As Schneider emphasizes, precisely modeling these phenomena is not just crucial. Still, it is the cornerstone of our work, enhancing the predictive fidelity of aerothermal simulations within the context of high-enthalpy hypersonic flows.

## 4. Discussion

The results of this computational study provide critical insights into the aerothermal behavior of a Mach 6 blunt cone and its associated boundary-layer transition phenomena. The simulation framework effectively captures shock wave formation, boundary-layer behavior, and surface heat transfer distribution, reinforcing the importance of incorporating high-fidelity numerical modeling techniques in hypersonic aerodynamics.

One of the primary observations in this study is the development of shock waves and their interaction with the boundary layer. Figure 6 highlights the fully developed shock wave structure, with the highest Mach number gradients occurring near the leading edge of the blunt cone. This behavior aligns with classical hypersonic flow theory, where the stagnation region experiences intense compression, leading to significant aerodynamic heating. The temperature distribution in Figure 7 further supports this observation, illustrating the highest thermal loads at the stagnation point, followed by a gradual decline along the cone's surface.

Despite the numerical approach's accuracy, discrepancies between computed heat transfer results and experimental findings from Stetson's 1983 study suggest the need for additional refinement in modeling techniques. The absence of a chemical kinetics model introduces limitations in accurately predicting non-equilibrium effects, particularly in high-enthalpy flows where vibrational relaxation and dissociation processes play a significant role. Prior studies, such as those by Schneider and Fedorov, emphasize that neglecting thermochemical non-equilibrium effects can lead to underestimation of heat transfer rates and inaccurate boundary-layer stability predictions.

Another key factor influencing the results is the surface roughness effects on boundary-layer transition. The findings indicate that discrete roughness elements, as modeled in the CFD simulations, contribute to early transition onset due to disturbance amplification in the boundary layer. This observation aligns with existing literature, which describes how surface discontinuities generate localized flow separations, leading to transition-inducing vortical structures. Future work should incorporate more detailed roughness modeling techniques, including stochastic distributions of roughness heights, to improve transition onset predictions.

The mesh independence study ensures that numerical diffusion and discretization errors are minimized, contributing to the stability and reliability of the computed results. The study employs structured grids with high orthogonality and low skewness, reinforcing the computational setup's robustness. However, additional refinement using adaptive mesh refinement (AMR) may be beneficial in further capturing shock-boundary layer interactions.

## 5. Limitations and Future Work

While this study successfully establishes a computational aerothermal framework for hypersonic boundary-layer transition analysis, some limitations remain. The one-way coupling approach used in the heat conduction solver may introduce small discrepancies in heat flux predictions, as it does not account for feedback interactions between fluid and solid domains. Future studies should explore a fully coupled CFD-thermal conduction solver to enhance predictive accuracy.

Additionally, incorporating Large-Eddy Simulation (LES) or Direct Numerical Simulation (DNS) approaches may offer higher-fidelity turbulence modeling, particularly for capturing small-scale instabilities within the boundary layer. The computational cost of such models remains a challenge, but hybrid RANS-LES models could balance computational efficiency and accuracy.

Lastly, experimental validation remains crucial for reinforcing computational findings. Future work should focus on wind and water tunnel testing or flight experiments to validate the numerical predictions presented in this study. Experimental validation can provide essential benchmark data for refining computational aerothermal models by integrating high-resolution Schlieren imaging and surface heat flux measurements.

## 6. Conclusion

In conclusion, this study underscores the critical importance of integrating advanced numerical techniques in hypersonic boundary-layer transition modeling. The findings elucidate key aerothermal phenomena, including shock wave development, surface heating, and boundary-layer instabilities, while identifying areas for further computational refinement. Future research should incorporate thermochemical non-equilibrium models to account for the complex chemical reactions and energy exchanges occurring at hypersonic speeds to enhance predictive accuracy. Implementing advanced turbulence solvers, such as Large-Eddy Simulation (LES) or Direct Numerical Simulation (DNS), can provide more detailed insights into the transitional and turbulent flow structures. Additionally, experimental validation efforts are essential to corroborate computational results and ensure the reliability of simulations. Collaborative approaches that combine high-fidelity simulations with empirical data will significantly advance the predictive capabilities of hypersonic aerothermal analyses.

By embracing these strategies, the aerospace community can achieve a more comprehensive understanding of hypersonic boundary-layer transitions, leading to improved design and performance of high-speed vehicles.

## 7. Future Research Directions and Literature Comparisons

1. **Advanced Modeling of Non-Equilibrium Effects**: Building upon our current two-temperature model, future studies will incorporate more comprehensive non-equilibrium models that account for additional energy modes and chemical reactions. This enhancement is expected to improve the accuracy of simulations in predicting aerothermal heating under hypersonic conditions.

2. **Experimental Validation**: To corroborate our computational findings, we plan to conduct controlled wind and water tunnel experiments replicating the conditions studied in our simulations. This empirical approach will provide valuable data to validate and refine our models, ensuring their applicability to real-world scenarios.

3. **Exploration of Active Flow Control Techniques**: Investigating active flow control methods, such as plasma actuation, offers potential for managing aerothermal heating in hypersonic flows. Recent studies have demonstrated the feasibility of plasma actuators in controlling hypersonic boundary layers, suggesting a promising area for further research.

4. **Investigation of Passive Control Strategies**: In addition to active methods, exploring passive control strategies, such as surface texturing or material selection, could provide insights into delaying or mitigating aerothermal heating. These approaches may offer practical solutions for hypersonic vehicle design without needing external energy input.

5. **High-Fidelity Simulations of Complex Geometries**: Extending our computational framework to more complex vehicle geometries will enhance the applicability of our findings. High-fidelity simulations that capture the intricacies of real-world designs are crucial for developing robust predictive tools for hypersonic boundary-layer behavior.

This study's computational aerothermal framework for analyzing hypersonic boundary-layer transition over a Mach 6 blunt cone aligns with and extends existing research in several key areas:

1. **Aerothermal Modeling Approaches:** This study uses high-fidelity numerical simulations consistent with contemporary efforts to model hypersonic boundary-layer transitions accurately. For instance, a study published in Advances in Aerodynamics utilized direct numerical simulation (DNS) to investigate hypersonic boundary-layer transitions, providing detailed insights into the transition mechanisms.

2. **Thermochemical Non-Equilibrium Effects:** The current study identifies the absence of thermochemical non-equilibrium modeling as a limitation, which is crucial for accurate heat transfer predictions in hypersonic flows. This observation is supported by findings in the literature, such as the analysis of thermochemical non-equilibrium hypersonic flow over a wave rider, which emphasizes the significant impact of non-equilibrium effects on aerothermal characteristics.

3. **Surface Roughness and Boundary-Layer Transition:** The influence of surface roughness on boundary-layer transition observed in this study aligns with existing research highlighting its substantial impact on aerothermodynamic loading. A report by NATO STO discusses how surface roughness can significantly affect the transition process, leading to increased heat transfer rates and altered boundary-layer behavior.

4. **Computational Frameworks and Validation:** The study's emphasis on mesh independence and structured grids with high orthogonality reflects best practices in computational aerothermal analysis. This approach aligns with methodologies discussed in the literature, such as developing geometry-independent hypersonic boundary-layer transition analysis frameworks, which stress the importance of robust computational setups for accurate simulations.

In summary, this study's aerothermal framework and analysis are consistent with current research trends, contributing valuable insights into hypersonic boundary-layer transition phenomena. Future work can further enhance the predictive capabilities of hypersonic aerothermal simulations by addressing identified limitations, such as incorporating thermochemical non-equilibrium models and advanced turbulence solvers.

**Author Contributions:** Conceptualization, V.K., and A.R.; methodology, A.R., C.D.C.; software, A.R., V.V.K. and P.K.; investigation, A.R., C.D.C.; resources, A.R.; writing—original draft preparation, A.R.; writing—review and editing, C.D.C., L.F.R., and R.O.A.; visualization, R.O.A..; supervision, V.K., and P.K.; project administration, A.R., and C.D.C.; funding acquisition, V.K. All authors have read and agreed to the published version of the manuscript.

**Funding:** The National Science Foundation Graduate Research Fellowship Program (NSF GRFP) under grant number 226101161A. The Air Force Office of Scientific Research funded this research under the Agile Science of Test and Evaluation (T&E) program grant number FA9550-19-1-0304. The U.S. Department of Energy Minority Serving Institutions Partnership Program (DOE-MSIPP) funded this research under grant number GRANT13584020.

**Data Availability Statement:** The data presented in this study are available from the corresponding author upon reasonable request.

**Acknowledgments:** I acknowledge Prof. Steven P. Schneider, Robert P. Velte, and Owen States from the School of Aeronautics and Astronautics at Purdue University for their guidance in this work.

**Conflicts of Interest:** The authors declare no conflicts of interest.

**Nomenclature:**

| | | | | | |
|---|---|---|---|---|---|
| $\rho$ | Density | $\left(\dfrac{Lbm}{in^3}\right)$ | $x$ | Spatial Dimension | $in$ |
| $u$ | Velocity | $\left(\dfrac{in}{s}\right)$ | $p$ | Pressure | $\left(\dfrac{lbs}{ft^2}\right)$ |
| $\tau$ | Shear Stress | $\left(\dfrac{lb}{in^2}\right)$ | $t$ | Time | $s$ |
| $e$ | Energy | $BTU$ | $\dot{q}$ | Heat Transfer Rate | $\left(\dfrac{BTU}{s}\right)$ |
| $T$ | Temperature | $R$ | $k$ | Conductivity | $\left(\dfrac{BTU}{in \cdot s \cdot F}\right)$ |
| $Cp$ | Heat Capacity | $\left(\dfrac{BTU}{Lbm \cdot F}\right)$ | $h$ | Heat Transfer Coefficient | $\left(\dfrac{BTU}{ft^2 \cdot s \cdot R}\right)$ |
| $F$ | Drag Force | $lbf$ | | | |